\documentclass[%
 sd,%
 amsmath,amssymb,
 reprint,%
 superscriptaddress, %
]{revtex4-1}

\usepackage{graphicx}
\usepackage{dcolumn}
\usepackage{bm}
\usepackage{makecell}
\usepackage{multirow} 
\usepackage{upgreek}

\usepackage{makecell}
\usepackage{multirow} 
\usepackage{hhline}
\usepackage{dcolumn}
\usepackage{ucs}
\usepackage{xcolor}

\newcommand{\degC}{\ensuremath{^\circ\textrm{C}}}
\newcommand{\capn}{\ensuremath{\mathrm{Ca}}}
\newcommand{\srm}[2]{\ensuremath{#1_\mathrm{#2}}}

\begin{document}

\preprint{AIP/123-QED}

\title{Influence of bidisperse self-assembled monolayer structure on the slip boundary condition of thin polymer films
}

\author{Joshua D. McGraw}
\email[]{joshua.mcgraw@phys.ens.fr}
\affiliation{Soft Matter Physics Group, Experimental Physics, Saarland University, 66041 Saarbr\"ucken, Germany}
 \affiliation{D\'epartement de Physique, Ecole Normale Sup\'erieure / PSL Research University, CNRS, 24 rue Lhomond, 75005 Paris, France} 
 
\author{Mischa Klos}
\affiliation{Soft Matter Physics Group, Experimental Physics, Saarland University, 66041 Saarbr\"ucken, Germany}

\author{Antoine Bridet}
\affiliation{Soft Matter Physics Group, Experimental Physics, Saarland University, 66041 Saarbr\"ucken, Germany}

\author{Hendrik H\"ahl}
\affiliation{Soft Matter Physics Group, Experimental Physics, Saarland University, 66041 Saarbr\"ucken, Germany}

\author{Michael Paulus}
\affiliation{Fakult\"at Physik/DELTA, TU Dortmund, 44221 Dortmund, Germany}

\author{Juan Manuel Castillo}
\affiliation{Laboratory of Engineering Thermodynamics, University of Kaiserslautern, Erwin-Schršdinger-Str. 44, 67663 Kaiserslautern, Germany}

\author{Martin Horsch}
\affiliation{Laboratory of Engineering Thermodynamics, University of Kaiserslautern, Erwin-Schršdinger-Str. 44, 67663 Kaiserslautern, Germany}

\author{Karin Jacobs}
\email[]{k.jacobs@physik.uni-saarland.de}
\affiliation{Soft Matter Physics Group, Experimental Physics, Saarland University, 66041 Saarbr\"ucken, Germany}
\date{\today}

\begin{abstract}
Alkylsilane self-assembled monolayers (SAMs) are often used as model substrates for their ease of preparation and hydrophobic properties. We have observed that these atomically smooth monolayers also provide a slip boundary condition for dewetting films composed of unentangled polymers. This slip length, an indirect measure of the friction between a given liquid and different solids, is switchable and can be increased [Fetzer \emph{et al.}, Phys. Rev. Lett. 2005; B\"aumchen \emph{et al.} J. Phys.: Cond. Matt. 2012] if the alkyl chain length is changed from 18 to 12 backbone carbons, for example. Typically, this change in boundary condition is affected in a quantized way, using one or the other alkyl chain length, thus obtaining one or the other slip length. Here, we present results in which this SAM structure is changed in a continuous way. We prepare SAMs containing bidisperse mixed SAMs of alkyl silanes, with the composition as a control parameter. We find that all the mixed SAMs we investigated show an enhanced slip boundary condition as compared to the single-component SAMs. The slip boundary condition is accessed using optical and atomic force microscopy, and we describe these observations in the context of X-ray reflectivity measurements. The slip length, varying over nearly two orders of magnitude, of identical polymer melts on chemically similar SAMs is found to correlate with the density of exposed alkyl chains. Our results demonstrate the importance of a well characterized solid/liquid pair, down to the angstrom level, when discussing friction between a liquid and a solid.  %
\end{abstract}

\maketitle

\section{\label{sec:level1}Introduction}

A molecule in a homogeneous bulk fluid is entirely surrounded by other molecules of the same type. Near a surface, however, some of the atoms or molecules surrounding a fluid molecule are necessarily different. In the context of polymer science, understanding of the bulk fluid is reaching maturity,~\cite{degennesscaling,doied, strobl96TXT,rubincolby} yet differences in the interactions between surface and fluid molecules can have profound effects on observed phenomenology which remain to be fully understood. Film structure effects include changes to chain packing and entanglement network statistics,~\cite{Silberberg1982, brownrussell, si05PRL, mcgraw13EPJE} chain end and short chain segregation,~\cite{hariharan93JCP,lee14PRL} van der Waals interactions between exposed and buried interfaces,~\cite{seeman01PRLb, lessel13PRL} and preparation effects resulting from non-equilibrium chain conformations.~\cite{Barbero2009, raegen10PRL, Clough2012, mcgraw13EPJE} Owing to possible melt-structure changes or interfacial physics, important dynamical examples of surface polymer phenomenology include: the apparent reduction in the glass transition temperature near a free interface~\cite{dalnoki01PRE,alcoutlabi05JPCM,paeng11JACS,baumchen12PRL,salez15PNAS}; and apparent viscosity reductions for entangled polymers.~\cite{Bodiguel2006, shinNATM07} As will be discussed in this article, a solid support may also promote slippage of a liquid polymer melt~\cite{degennes85RMP, degenneslecture, reiter00PRL, leger03JPCM, priezjev04PRL, fetzer07LMR_thr, baumchen12JCM, mcgraw14JCIS, tretyakov13SM, chenneviere16MAC} which is flowing near the solid/liquid boundary. The microscopic details of this boundary play a huge role in the amount of slippage that is observed. 

The solid/liquid boundary condition is normally taken to be that of a no-slip type, with no flow relative to the solid boundary. At microscopic scales, however, this empirical boundary condition may indeed fail;~\cite{neto05RPP, lauga07TXT, bocquetSCR10} there is no general physical reasoning that the relative flow velocity at the solid/liquid boundary must be zero. The linear Navier slip model~\cite{NavierOLD} writes the velocity at the boundary in terms of a stress balance at the interface
\begin{align}
	\kappa v_{x} = \eta\partial_{z} v_{x}\ , z=0\ ,
\label{slipEq}
\end{align}
where $v_x$ is the fluid velocity parallel to the substrate, $\partial_z$ denotes differentiation normal to the interface, $\eta$ is the viscosity and $\kappa$ is the linear friction coefficient. Using such a stress balance allows one to define the slip length, $b$, as a linear extrapolation of the velocity profile through two material parameters, $b = \eta/\kappa$; this relation indicates that the slip length is inversely related to the linear friction coefficient between the solid and the liquid. While small molecule and unentangled polymer slip lengths are normally limited to some 10's of nanometres,~\cite{lauga07TXT} polymer slippage can be much larger, up to 10$\,\mu$m~\cite{reiter00PRL, leger03JPCM, baeumchen09PRL} for entangled polymers on polymer brushes or smooth hydrophobic substrates. 

\begin{figure}[b!]
\includegraphics[width=3.4in]{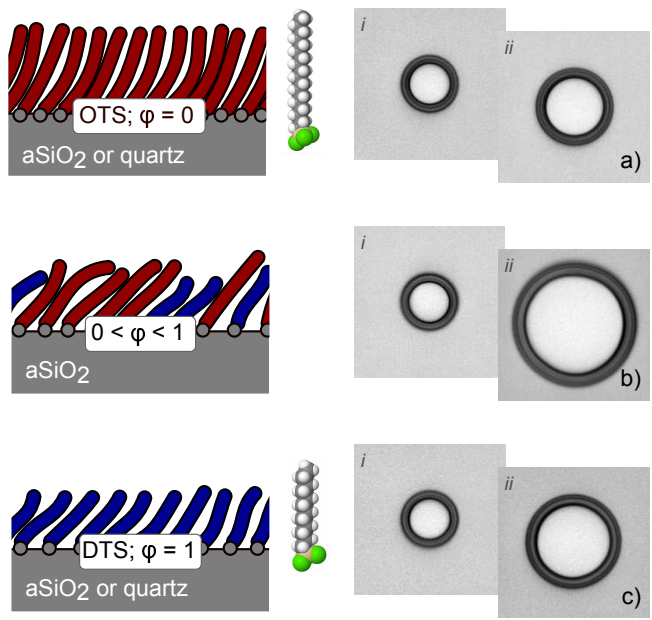}
\caption{\label{schemopt} Left column: schematics of the SAMs used in this study; in a)-c) pure OTS; mixed SAMs prepared with DTS volume fraction, $\phi$; and pure DTS monolayers are shown, respectively. The red molecules represent OTS, shown to the right of the schematic in a) (hydrogen, white; carbon, grey; chlorine, green; silicon, flesh). Blue molecules represent DTS, shown in c). OTS layers are {\em ca.} 2\,nm thick, and in the case of native aSiO$_2$ substrates, a 500\,$\mu$m thick Si layer (not shown) exists as a bottom layer. Right column: optical micrographs of PS films with thickness $h_0 = 160\pm4$\,nm dewetting from the SAMs at 110\,\degC: a) pure OTS, b) $\phi = 0.75$, c) pure DTS. In images $i$, the radii are $R=4.9\pm0.1\,\mu$m, and the following images $ii$ were taken after a constant time interval, $t_{ii}-t_i = 18$\,min. }
\end{figure}

One method of preparing hydrophobic substrates is through the use of self-assembled monolayers~\cite{sagiv80JACS, wasserman90PRB, tidswell90PRB, *tidswell91JCP, lestelius99CnSB, schreiber00PSS, love05CR, genzer08LAN, gutfreund13PRE, lessel15LAN} (SAMs), which are schematically depicted in Figure.~\ref{schemopt}. Properly prepared SAMs provide an easy route to making highly uniform substrates with, for example, controlled wettability~\cite{genzer08LAN, lestelius99CnSB} and other~\cite{schreiber00PSS, love05CR, genzer08LAN} properties. A particular property that has had significant attention is solid/solid friction~\cite{booth09LAN,barrena99PRL,chandross04PRL,vilt09JPCC} on alkylsilane and alkanethiol SAMs. Collectively, these studies show that the specific structure of the SAM also plays a large role in the observed friction, with subtle changes in the structure of the monolayer leading to observable changes in solid/solid friction. In this article, we extend these observations to liquid/solid friction, continuously varying the structure of SAMs by preparing mixed alkylsilane monolayers. In these experiments, the liquid ({\em i.e.} the frictional probe) is always identical, yet the slip length on the chemically identical SAMs with differing structure changes by orders of magnitude. 

The paper is organized in the following way: in Section~\ref{sec:expts} we outline the sample preparation and provide a description of the dewetting and X-ray reflectivity (XRR) experiments. Section~\ref{slipdet} describes the data processing for slip length determination and Section~\ref{rhodet} describes the XRR data processing which leads to the electron density profiles. In Sections~\ref{slipdesc} and~\ref{nanodesc} we provide an analysis of the slip length and substrate nanostructure, respectively, and in Section~\ref{slipnanodesc} we provide a prospective link between the nanostructure and slip length determination. Following a conclusion, we provide four appendices containing additional dynamical dewetting data (Appendix~\ref{appA}), supplementary quantities derived from the electron density profiles~(\ref{appB}), molecular dynamics results~(\ref{appC}), and a table of all X-ray reflectivity fitting parameters~(\ref{appD}).

\section{\label{sec:expts} Experiments}

As substrates, Si wafers (crystal orientation (100); Si-Mat Silicon Materials, Kaufering, Germany, roughness of \emph{ca.} $0.2$\,nm on $1\,\mu$m$^2$ area determined using atomic force microscopy (AFM)) with a native, amorphous SiO$_2$ were used. We denote these substrates as aSiO$_2$. Crystalline quartz wafers with no amorphous SiO$_2$ surface layer were also used, and are referred to as xSiO$_2$. These xSiO$_2$ wafers (34$\,^\circ$ cut angle relative to the $xy$-plane, Nano Quarz Wafer GmbH, Langenzenn, Germany) were polished on both sides by the manufacturer resulting in translucent appearance with a roughness below 0.5 nm. This double-sided polishing minimizes deflections of the wafer making it possible to obtain XRR data of comparable quality to that obtained from aSiO$_2$ wafers. 

Hydrophobization of the substrates was accomplished via the deposition silane self-assembled monolayers (SAMs); as silanes, we used 12-C and 18-C, dodecyltrichlorosilane and octadecyltrichlorosilane (DTS and OTS, respectively; purchased from Sigma-Aldrich, Steinheim, Germany). The recipe detailed in ref.~\onlinecite{lessel15LAN} was used throughout. In order to produce the mixed monolayers, we prepared the usual liquid phase deposition solutions, but added silane volume fractions, $\phi$, of DTS and silane volume fractions, $1-\phi$, of OTS. All other steps, including wafer cleaning, deposition and rinsing, were followed as usual. On xSiO$_2$ wafers, only pure monolayers of OTS and DTS were prepared. 

As a liquid, polystyrene (PS) with weight averaged molecular weights 10.3 kg/mol (used with aSiO$_2$ substrates) or 12.5 kg/mol (used with aSiO$_2$ and xSiO$_2$ substrates), both with polydispersity index 1.03, were used (Polymer Standards Service, Mainz, Germany). The polymer was dissolved in toluene (chromatography grade; Merck, Darmstadt, Germany) in varying concentrations which, when spincoated onto freshly cleaved mica sheets (grade V2; Plano GmbH, Wetzlar, Germany) yielded thicknesses on the range of $110 \lesssim h_0 \lesssim 200$\,nm. After spin coating, the films were floated onto the surface of an ultra clean water bath (18 M$\Omega$\,cm, total organic carbon content $<$~6\,ppb; TKA-GenPure, ThermoFisher, Darmstadt, Germany) and picked up from the surface using hydrophobized substrates. Before floating, the substrates were sonicated in ethanol, acetone and toluene (all chromatography grade; Sigma-Aldrich, Steinheim, Germany) for three to five minutes and dried using dry N$_2$ following each sonication; film deposition occurred immediately after drying. 
\begin{figure}[t!]
\includegraphics[width=3.4in]{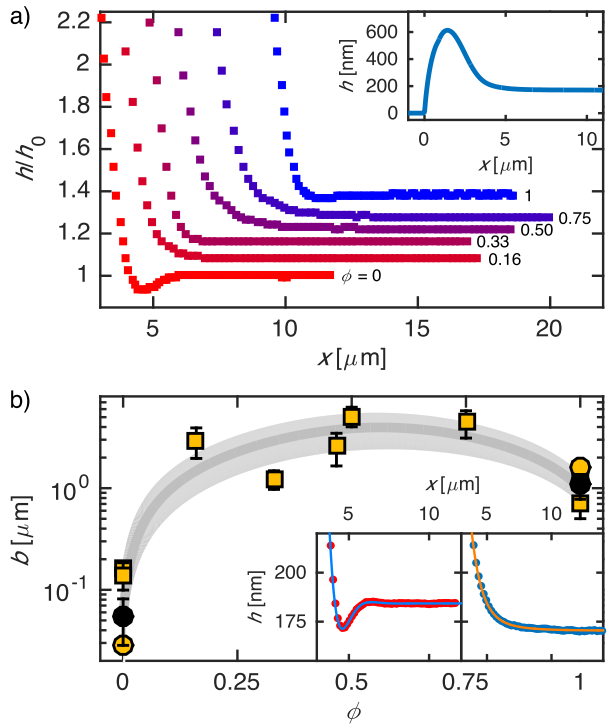}
\caption{\label{slipfig} a) AFM data showing the height profiles on the `wet side' of the dewetting 10.3 kg/mol PS rim. DTS volume fraction, $\phi$, is shown next to the curves, and the curves are shifted horizontally and vertically for clarity. Starting from $\phi = 0$ and ascending, $h_0$ for the films were 184, 184, 181, 170, 200, and 165\,nm, each with an error of 1\,nm. The inset shows a full height profile of a film dewetting from a SAM ($\phi=0.5$) with the substrate ($x<0$), undisturbed PS film ($x\gg5\,\mu$m) and the rim ($0<x\lesssim5\,\mu$m). b) Slip length as a function of DTS volume fraction for aSiO$_2$ (yellow) and xSiO$_2$ (black); squares show data for 10.3 kg/mol PS, circles show 12.3 kg/mol PS data. The grey region is a guide for the eye. Each data point represents an independent silanization, and the error bars represent the statistical deviation over several measurements of $b$ comprising measurements from different holes in different dewetting films; insets show fits for an oscillatory profile (OTS, left) and a monotonic profile ($\phi = 0.5$, right). }
\end{figure}

After preparation of the thin film samples, the wafers were placed onto a Linkam hot stage at 110\,\degC\ under optical microscopy (Leitz). Isolated holes were nucleated randomly in the film and images as shown in Figure~\ref{schemopt}b) were taken periodically (typically 10\,s to 1\,min between frames). When the holes reached 12\,$\mu$m in radius, samples were quenched to room temperature within seconds. The same 12\,$\mu$m holes that had been imaged under optical microscopy were subsequently investigated using atomic force microscopy (AFM; Dimension Fastscan or Multimode, Bruker, Karlsruhe, Germany). Rim profiles were collected for several such holes at each composition and for each substrate used, examples are shown in Figure~\ref{slipfig}. 
\begin{figure}[t!]
\includegraphics[width=3.4in]{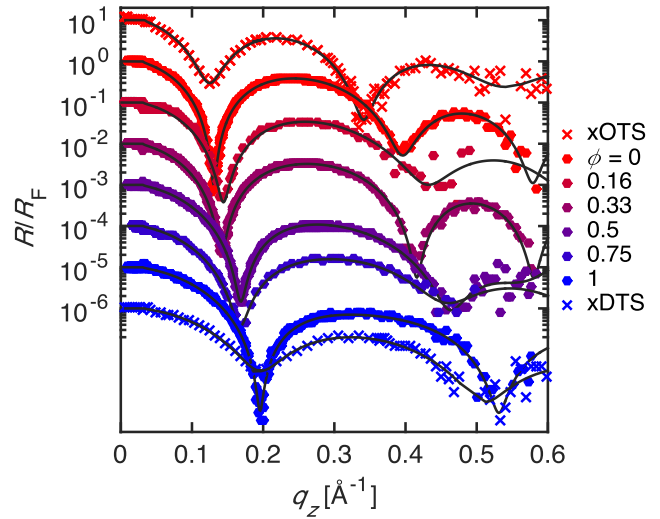}
\caption{\label{XRRraw} X-ray reflectivity, $R$, normalized by the Fresnel reflectivity, $R_\textrm{F}$. Data from SAMs on xSiO$_2$ are denoted in the legend with x's (\emph{i.e.} xOTS and xDTS). The remaining curves show the series of OTS/DTS mixed SAMs $0\leq\phi\leq1$ on aSiO$_2$. Solid black curves are model reflectivities described in the text and shown in Figure~\ref{EDPmods}. For clarity, the curves (except that for OTS on aSiO$_2$) are shifted vertically from one another by one decade. }
\end{figure}

To characterize the silane layers, bare, cleaned SAM-coated substrates were examined with X-ray reflectivity (XRR) at beamline BL9 of the synchrotron light source DELTA (Dortmund, Germany).~\cite{krywka2006} A photon wavelength of $\lambda = 0.459$\,\AA\ was used to investigate the monolayers in $\theta$ - $2\theta$ geometry. For the xSiO2 substrates and for $\phi = 0.33$ and $\phi = 0.75$, a Bruker D8 ADVANCE diffractometer was used, operating in $\theta$ - $\theta$ geometry with a copper anode (Cu K-$\alpha$, wavelength  $\lambda = 1.54$\,\AA) and a G\"obel mirror for beam-parallelization. Figure 3 shows the reflectivity curves, normalized by the Fresnel reflectivity~\cite{tolan99TXT} (\emph{i.e.} the reflectivity of a pure, smooth, single-component substrate) of Si or xSiO$_2$.

\section{Data analysis}

\subsection{\label{slipdet}Slip length}
\begin{figure*}
\centering
\includegraphics[width=\textwidth]{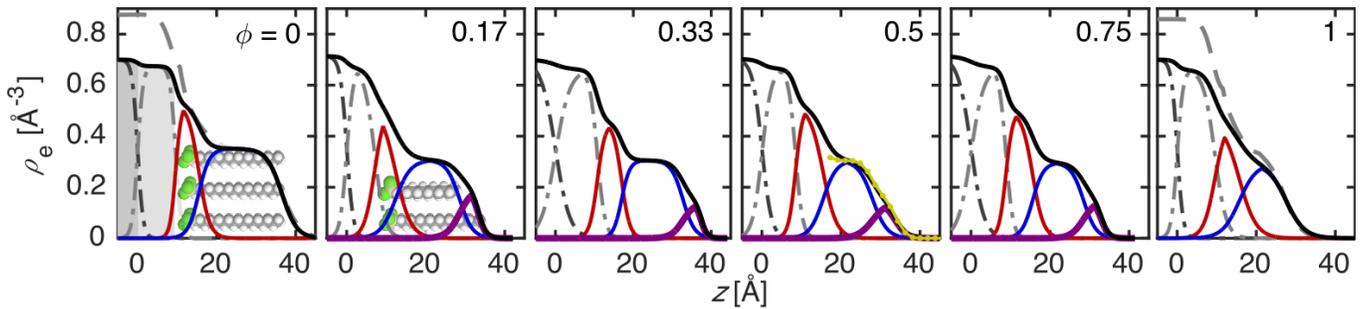}
\caption{\label{EDPmods} Electron density, $\rho_\textrm{e}$, (heavy black lines) as a function of depth, $z$, for all the SAMs on SiO$_2$. The total electron density for the SAMs on xSiO$_2$ are shown as grey dashed lines in the panels for $\phi = \{0,1\}$. For all compositions on Si, $z=0$ is taken as the point of inflection in \srm{\rho}{e}\ between Si and SiO$_2$ (thin black and grey dash-dotted lines, respectively). Also shown are the head group (red), tail group (blue), and for $\phi\neq\{0,1\}$, the second tail group (violet). A schematic of the physical layer structure is shown underlying the data in the $\phi=0$ panel; OTS (lower) and DTS (upper) molecules are schematically shown to scale underlying the data in the panel for $\phi = 0.17$. In the panel for $\phi = 0.5$, the yellow dotted line represents data obtained from MD simulations with $\srm{\sigma}{C} = 4.5$, and $\phi = 0.5$ with completely uncorrelated placement of OTS and DTS molecules. For clarity, only the portion overlapping with the tail layer is shown. 
}
\end{figure*}

The evaluation of optical and atomic force microscopy data to determine slip lengths is well documented elsewhere,~\cite{fetzer07LMR_thr, baumchen12JCM, mcgraw14JCIS} yet for completeness we briefly summarize the method here. The most accurate values of slip length determined from dewetting experiments are obtained by analyzing the decay of dewetting rims into the undisturbed film. Such decays are shown in Figure~\ref{slipfig}a), and can be seen to take forms that are either monotonic or oscillatory. These data may be fit to equations of the form 
\begin{subequations}
\begin{align}
	h(x) &= h_0 + A_1e^{-k_1x} + A_2e^{-k_2x}\qquad \mathrm{[monotonic}]\ , \\
		h(x) &= h_0 + Ae^{-\srm{k}{r}x}\cos(\srm{k}{i}x+\varphi)\qquad\mathrm{[oscillatory}]\ , 
\end{align} 
\label{monoosc}
\end{subequations}
where $x$ is the position relative to the contact line, $\varphi$ is a phase shift, $A$ are amplitudes, and the various $k$ are either inverse decay lengths or wavenumbers. With $k$ obtained from fits to the profiles (insets of Figure~\ref{slipfig}b)), it is possible to solve the characteristic equation, obtained assuming a flat rigid substrate and a purely viscous fluid,~\cite{fetzer07LMR_thr, baumchen12JCM, mcgraw14JCIS} 
\begin{align}
	\left(1+\frac{h_0}{3b} \right)(h_0k)^3 + 4\capn\left(1+\frac{h_0}{2b} \right)(h_0k)^2  - \capn\frac{h_0}{b} &= 0\ , 
	\label{chareq}
\end{align}
to obtain the capillary number, $\capn = \eta \dot{R}/\gamma$, and the slip length, $b$. In \capn, $\gamma$ is the surface tension taken as~\cite{wu70JPC} 30.2 mJ/m$^2$ and $\dot{R}$ is the contact line velocity at the moment of quenching, see Figure~\ref{fulldyn}. In a previous study~\cite{mcgraw14JCIS} using the same polymer at the same temperature under the same microscope as used here, the viscosity was determined to be $\eta = (1.4\pm1)$\,MPa\,s$^{-1}$. Through the capillary number, this value of the viscosity was used as an independent check on the slip length obtained from the present fits; the velocity of the contact line $\dot{R}$ is known from optical measurements (Figure~\ref{fulldyn}). We note lastly that in cases for which the profile is monotonic, it may happen that a single exponential, with a single decay length $k$, adequately describes the data. In such cases, knowledge of the viscosity, surface tension, and contact line velocity when the polymer melt was quenched, was used to define the capillary number, and therefore extract a slip length from Equation~\ref{chareq} ($h_0$ is also obtained from the AFM measurement). 

\subsection{\label{rhodet}Electron density}

Using the effective density model,~\cite{tolan99TXT} electron density
profiles (EDPs), $\srm{\rho}{e}(z)$, were constructed as a collection of slabs with thickness $d$, interfacial roughness $\sigma$ and electron density $\rho_\textrm{e}$. The chosen model ensures continuous profiles even for high interfacial roughness.

For single-component SAMs, models incorporating slabs for Si (semi-infinite), aSiO$_2$, silane head group, silane tails, and air (semi-infinite) were used. SAMs on xSiO$_2$ were modelled with semi-infinite xSiO$_2$ and air lay- ers, with the silane heads and tails comprising the middle layers. For mixed monolayers, an additional tail layer~\cite{lestelius99CnSB} was used.

The profile is sliced into layers with constant electron density and a thickness of less than 0.1\,\AA. This stack of thin layers is used as an optical model to compute a theoretical reflectivity using the matrix formalism of the dynamical scattering theory.~\cite{daillant09TXT} To fit the experimental reflectivity, the model refinement was in turn performed using home made software on \textsc{Matlab}. A detailed accounting of the fitting parameters is shown in Table~\ref{XRRtable}, and the EDPs as a function of depth are shown in Figure~\ref{EDPmods}.

\section{Results and Discussion}

\subsection{\label{slipdesc}Dewetting and slip length}

The right column of Figure~\ref{schemopt} shows optical microscopy images of holes growing in PS films on three different SAMs. In these images, the substrate (bright center) is separated from the undisturbed film (grey) by the rim collecting dewetted PS (dark ring, {\em cf.} the height profile in the inset of Figure~\ref{slipfig}). Images {\em i} are chosen such that the holes have identical radii, $R = 4.9\pm0.1\,\mu$m, and images {\em ii} are chosen such that $t_{ii}-t_i = 18$\,min elapses between series {\em i} and series {\em ii}. Clearly, the film on the mixed SAM has the fastest hole growth for this particular comparison. 

We have additionally performed detailed hole growth investigations for all the compositions, $\phi$, used here. The full hole growth dynamics, $R(t)$, are shown for an exemplary series of films with identical thickness in Appendix~\ref{appA}, Figure~\ref{fulldyn}. In addition to an observed reproducibility for the dynamics between independent films on similar substrates, in this figure it is also seen that holes in PS films grow the slowest on SAMs composed of OTS only. The next slowest dynamics are seen on pure DTS SAMs. All of the mixed SAMs promote enhanced hole growth dynamics as compared to the single-component SAMs, with the fastest dynamics observed for $\phi = 0.5$. While we show data for only one film thickness in Figure~\ref{fulldyn}, we note that the same trends in the dynamics are seen as well for films with thicknesses $h_0 \approx 110$\,nm and $h_0 \approx 180$\,nm. 

Given the different dynamics presented by the different SAMs, we first investigate the receding contact angles, $\theta$, of PS(10k) to give an indication of the driving force, through the spreading coefficient~\cite{degennes03TXT} $S=\gamma(\cos\theta-1)$. Contact angles were measured from sessile droplets of polymer formed after complete dewetting of a film and 1\,h of equilibration at 140\,\degC. Figure~\ref{angles_sigma}a) shows the contact angle of PS, $\theta$, as a function of composition. There it is seen that the contact angles of PS on pure SAMs are largest, and within error, identical.~\cite{baumchen12JCM,mcgraw14JCIS} Consistent with the trends shown in gradient deposition of alkanethiols on gold,~\cite{lestelius99CnSB} all of the mixed-component SAMs studied here present a smaller receding contact angle as compared to the contact angles on pure SAMs. In the context of dewetting thin liquid films, it is well established~\cite{redon91PRL} that for no-slip dewetting, the expected dewetting velocity is given by~\footnote{The quoted cubic dependence on contact angle of the dewetting velocity is strictly valid for small angles. Although not necessarily the case for the experiments here, the dewetting velocity is always increasing~\cite{bonn09RMP} with $\theta$, and is opposite to the trend as seen here.} $\dot{R} = \gamma\eta^{-1}\theta^3$. Thus, our results are completely at odds with the no-slip theory since a higher contact angle is expected to yield a higher dewetting rate. This fact alone is not striking, since previous results,~\cite{fetzer07LMR_thr, baumchen12JCM, mcgraw14JCIS} along with those presented here for the pure components, show different dewetting rates for PS on OTS and PS on DTS, with nearly identical contact angles. This difference in the dewetting rates is attributed~\cite{fetzer07LMR_thr, baumchen12JCM, mcgraw14JCIS} to the presence of a slip boundary condition at the solid/liquid interface. 

In Figure~\ref{slipfig}a) height profiles of dewetting PS films for SAMs with $0\leq\phi\leq1$ on aSiO$_2$ are shown. As described in ref.~\onlinecite{fetzer07LMR_thr}, a profile with a slowly decaying, monotonic rim is characteristic of a larger slip length as compared to a profile with a rim that has an oscillatory shape, provided the dewetting films are the same. Consistent with the dynamics of hole growth ({\em cf.} Figures~\ref{schemopt} and~\ref{fulldyn}), the monotonic rim profiles for mixed monolayers suggest a higher slip length than those oscillatory decays observed for the pure monolayers in Figure~\ref{slipfig}a). With a full analysis of the rim profiles according to Equations~\ref{monoosc} and~\ref{chareq}, in Figure~\ref{slipfig}b) we show the slip length as a function of the DTS volume fraction for each of the monolayers prepared. We find that the slip length is nonmonotonic with composition, reaching a maximum for $\phi = 0.5$. Specifically, the slip length is $5\pm1\,\mu$m, which is approximately a factor of 50 larger than the slip length presented by the OTS, with $b_\textrm{OTS} \sim 100$\,nm. 

To check for possible influences of lower substrate layers beyond the SAM, we prepared pure OTS and DTS monlayers on xSiO$_2$. The slip lengths measured using xSiO$_2$ as a base substrate are consistent with those on SAMs using aSiO$_2$ as a base; we thus conclude that the liquid/solid friction is dominantly governed by the SAM. In order to explain the observations made in this section, we now move on to an investigation of the structural features of the SAMs. 

\subsection{\label{nanodesc}Substrate nanostructure}

To investigate the role of substrate nanostructure on the slip length, we have used XRR to characterize the electron density profiles. In Figure~\ref{EDPmods} are shown the model volumetric electron densities~\cite{lessel15LAN, gutfreund13PRE, tidswell90PRB, *tidswell91JCP} that were used to reproduce the reflectivities shown in Figure~\ref{XRRraw}. For the cases $\phi = \{0,1\}$, the EDPs for monolayers on aSiO$_2$ and xSiO$_2$ are equivalent within the expected deviation between silanizations on different aSiO$_2$ wafers (\emph{cf.} also refs.~\onlinecite{lessel15LAN, gutfreund13PRE} for OTS and DTS on aSiO$_2$); these EDPs are furthermore consistent with molecular dynamics simulations of SAMs on cristobalite substrates.~\cite{castillo15LAN} The equivalence of the bare SAM structure corroborates the equivalence of PS slip lengths on pure SAMs grafted to aSiO$_2$ and to xSiO$_2$. Therefore, we can conclude that the slip length of a liquid/solid pair is dominantly controlled by the structure of the SAM with which it is in contact, notwithstanding corrections that may arise from compositions deeper within the substrate.~\cite{lessel13PRL} 
\begin{figure}[b!]
\includegraphics{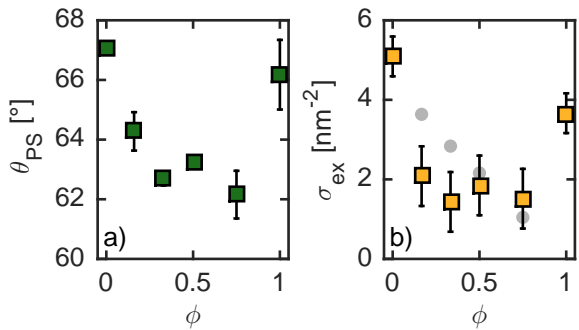}
\caption{\label{angles_sigma} a) Receding contact angles of PS on SAMs with various concentrations of DTS. Errors represent statistical uncertainty over several measured droplets. b) Exposed chain density, $\sigma_\textrm{ex}$ (see Equation~\ref{exchainsmain}) as a function of composition, $\phi$ for the SAMs investigated using XRR as in Figure~\ref{EDPmods}. The gray circles are an estimate assuming that the volume fraction of DTS in the monolayer is equivalent to the volume fraction of DTS in the preparation solution. }
\end{figure}

Moving onto an analysis of the mixed monolayers, the electron density profiles were modelled as for the pure monolayers, additionally decomposing the outermost (alkyl tail) layer into two sublayers. Using this model, we found better fits to the raw data presented in Figure~\ref{XRRraw}, with a typical reduction in the goodness of fit by a factor of 30\% or more. The model is motivated furthermore by experimental observations~\cite{lestelius99CnSB} and molecular dynamics simulations of mixed monolayers. This second tail layer accounts for the fact that OTS molecules are longer than DTS molecules, as shown schematically in Figures~\ref{schemopt} and~\ref{EDPmods} (see panels $\phi = 0$ and $\phi = 0.17$). Since the molecules are tightly packed (discussed in the following paragraph), we expect that the OTS chain end will tend to be found nearer to the air/SAM interface; this hypothesis is confirmed by MD simulations of the type performed in ref.~\onlinecite{castillo15LAN}, see Figure~\ref{MDfig}. 

The layer thickness and grafting density are interrelated quantities that result from~\cite{schreiber00PSS} a balance between i) the binding strength of a terminal Si atom and the SiO$_2$ substrate as well as the silane cross linking network~\cite{sagiv80JACS} on one hand and ii) a propensity for the molecules to rearrange between binding sites and within the network, as dictated by entropic mixing and van der Waals attractions between individual silane molecules. As seen previously~\cite{lessel15LAN, gutfreund13PRE} and confirmed here, the net result for our pure systems is that OTS packs more tightly ($\srm{\sigma}{C} \approx 5.1\,\textrm{nm}^{-2}$) than the typical DTS packing ($\srm{\sigma}{C} \approx 3.7\,\textrm{nm}^{-2}$), see Appendix~\ref{appB} for details, Equations~\ref{electrondensity} and~\ref{graftedchains}. These packing densities correspond well with the typical densities of crystalline $n$-alkanes~\cite{muller48, craig94JMC}, which for $n=18$ promote an areal density of $\srm{\sigma}{C} \approx 4.82$\,nm$^{-2}$. The total SAM alkylsilane tail layer thicknesses, $d$, and grafting density, \srm{\sigma}{C}, are shown as a function of composition in Appendix~\ref{appB}, Figure~\ref{tailschains}. The grafting densities of all the mixed SAMs lie between those of the extremal \srm{\sigma}{C}\ of the pure components; the total tail layer thickness also smoothly decreases with composition. Thus, both $d$ and \srm{\sigma}{C}\ are monotonic with composition and we do not expect either of these quantities to independently control the slip length, which is nonmonotonic with composition, Figure~\ref{slipfig}b). 

We now make an estimate for the density of alkylsilane chains that are exposed, \srm{\sigma}{ex}, at the air/SAM interface. For the pure SAMs, we assume equivalence between \srm{\sigma}{C}\ and \srm{\sigma}{ex}. For the mixed SAMs, a simple estimate for the exposed chain density is made by integrating the EDP for the second tail layer, $\sigma_2 = \int dz\ \rho_2(z)$, (violet in Figure~\ref{EDPmods}, `Tail 2' in Table~\ref{XRRtable}) and comparing this to the same integration, $\sigma_1 = \int dz\ \rho_1(z)$, of the first tail layer (blue, `Tail 1'). We assume that $\sigma_2$ is associated solely with the ends of OTS chains that extend beyond a DTS underlayer ({\em i.e.}, a tight packing assumption confirmed by the MD results, see Figure~\ref{MDfig}). Accounting for the electrons associated with the C and H atoms in the molecules, the average number of electrons associated with partial molecules in this outer region is $\srm{N}{out} = \srm{N}{OTS}-\srm{N}{DTS} = 48$. Thus, the exposed chain density is
\begin{align}
	\srm{\sigma}{ex} = 
	\begin{cases}
	    \srm{\sigma}{C}\ ,			& \text{if } \phi = \{0,1\}\ ,\\
	    \srm{\sigma}{2}/\srm{N}{out}\ , 	& 0<\phi<1\ .
	\end{cases}
\label{exchainsmain}
\end{align} 
In Figure~\ref{angles_sigma} is shown \srm{\sigma}{ex} as a function of $\phi$. For comparison, an estimate assuming that $\phi$ in the preparation solution is the same as that in the resulting SAM is given. The general nonmonotonic trend with $\phi$ is common between these two estimates; see Appendix~\ref{appB} for details. 

\subsection{\label{slipnanodesc}Nanostructure and slip}

In Figure~\ref{angles_sigma}b) we see that \srm{\sigma}{ex}\ shows a nonmonotonic trend with DTS volume fraction, with a minimum at a volume fraction $\srm{\phi}{min} \approx 0.5\pm 0.2$. In the context of polymer slip, the exposed chain density may have an impact on the friction resulting from several possible mechanisms. While the available data cannot distinguish between these mechanisms, we briefly describe them in the following. 

In the context of velocity dependent rubber/solid friction,~\cite{vorvolakos03LAN, shallamach63WR, chernyak86WR} the measured frictional stress is typically written as $\srm{\tau}{fric} \sim \srm{\sigma}{chain}$ where \srm{\sigma}{chain}\ is the density of polymer chains in contact with the solid; these being the objects which carry the frictional load. Similarly, if we assume that the exposed alkyl chains are dominantly responsible for transferring the frictional load from the sliding polymer melt down to the rest of the solid substrate ({\em i.e.} the aSiO$_2$ and Si), fewer direct contacts with the substrate leads to a weaker friction. The slip lengths (inversely proportional to the solid/liquid linear friction coefficient, see Equation~\ref{slipEq}) of Figure~\ref{slipfig}b) follow inversely the trend of exposed chain density with composition. 

In addition to the number of direct contacts with the underlying substrate, also the nature of the contacts may play a role in the friction. The lower total grafting density presented by the DTS SAMs as compared to the OTS SAMs causes the DTS chains to lie flatter against the underlying substrate.~\cite{lessel15LAN, gutfreund13PRE, castillo15LAN} Likewise, the exposed OTS chains in the mixed monolayers take on configurations with an enhanced proportion of gauche conformations as compared to pure monolayers, as evidenced experimentally by Lestelius and coworkers~\cite{lestelius99CnSB} and also suggested by Vilt and coworkers.~\cite{vilt09JPCC} An enhancement of CH$_2$ groups at the surface is furthermore indicated in the MD simulations by the relatively lower probability of finding terminal CH$_3$ groups near the interface in mixed SAMs as compared to pure SAMs (Figure~\ref{MDfig}). The presence of gauche conformations at the surface, along with flatter-lying alkyl chains promotes surface exposure of nonterminal CH$_2$ groups. To the extent that the energetic interaction between such CH$_2$ groups and the phenyl rings of PS is different,~\cite{bocquet94PRE, bocquetSCR10, huang14PRE} the PS/SAM friction may also be different. 

Lastly, lateral structure of the atomically smooth SAMs (RMS roughness {\em ca.} 0.2 nm using AFM) may play a role in substrate friction. While we have been unable to detect such lateral differences, {\em e.g.} phase separation,~\cite{schreiber00PSS, lestelius99CnSB} using tapping mode AFM down to sub-nanometre pixel sizes, the sub-nanometric differences in the depth profiles of electron density may be partly due to different lateral roughness profiles at the atomic scale.~\cite{tolan99TXT} To the extent that this atomic scale roughness may influence the packing of PS chains and their side groups next to the substrates, as evidenced by XRR~\cite{gutfreund13PRE} and sum frequency generation with sapphire substrates,~\cite{gautam00PRL} the friction may be impacted. 

In the previous paragraphs, we have outlined possible mechanisms for the slip enhancement on mixed SAMs as compared to pure SAMs. Whatever the dominating mechanism or combination of mechanisms, we see that subtle changes in the SAM structure, as revealed here by XRR and MD, lead to a nearly two orders of magnitude difference in the slip length experienced between an identical liquid and the various solid substrates used. 

\section*{conclusion}

We have studied unentangled, liquid polymer films dewetting from pure and mixed self-assembled monolayers on SiO$_2$ substrates. The SAMs were composed of fully hydrogenated C$_{18}$ (OTS) and C$_{12}$ (DTS) alkylsilanes, with various compositions, $0\leq\phi\leq1$, of DTS in the monolayers. Observation of the dewetting rim profiles by AFM, and the dewetting dynamics using optical microscopy, gives access to the slip length of the polystyrene/solid pairs. By preparing mixed SAMs, we have observed a 5-fold increase in the slip length at a composition of $\phi = 0.5$ compared to pure DTS. More dramatically, the slip length of PS on pure OTS is 50 times smaller than that of the monolayer prepared at $\phi = 0.5$; all intermediate compositions give enhanced slip as compared to the pure SAMs. These large changes in the slip length have been discussed in the context of electron density profile measurements accessed through X-ray reflectivity. We find that the exposed monolayer chain density correlates well with the observed non-monotonic dependence of the slip length with composition, and propose mechanisms for the observed dependence. Our study demonstrates the need for detailed structural knowledge of a SAM surface in order to understand liquid friction on solid supports.

\begin{acknowledgments}
The authors gratefully acknowledge NSERC of Canada and the German Science Foundation for financial support. JDM was supported by LabEX ENS-ICFP: ANR-10-LABX-0010/ANR-10-IDEX-0001-02 PSL. I. Kiesel at TU-Dortmund, DELTA is gratefully acknowledged for technical support for the XRR measurements. M. Chaudhury and R. Ledesma are warmly thanked for useful discussions. 
\end{acknowledgments}

\appendix
\renewcommand\thefigure{\thesection.\arabic{figure}}  
\setcounter{figure}{0}

\section{\label{appA}Hole growth dynamics}
\begin{figure}[b!]
\includegraphics{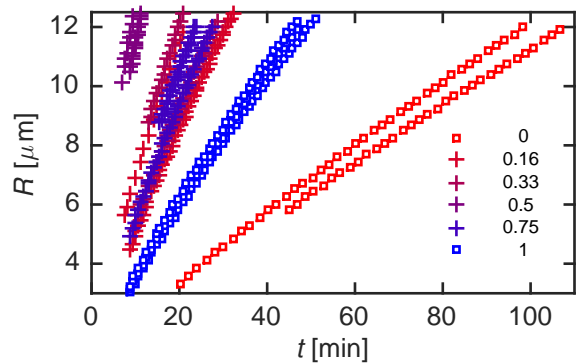}
\caption{\label{fulldyn} Hole radius, $R$, as a function of time, $t$, for 10 kg/mol PS films dewetting at 110\,$^\circ$C with thicknesses $h_0 = 160\pm4$\,nm. Radii are obtained from optical data as in Figure~\ref{schemopt}. Each data set is taken from separate hole growth measurements from different samples, demonstrating the reproducibility of the results. The legend indicates the composition, $\phi$, of DTS in the substrate SAM. }
\end{figure}

The right column of Figure~\ref{schemopt} shows optical microscopy images of hole growth in PS films. Here we present full dynamics $R(t)$ for several holes over all compositions. Shown in Figure~\ref{fulldyn} are the dynamics for $0\leq\phi\leq1$ for at least two holes growing in films with thicknesses $h_0 = 160\pm4$\,nm. The dynamics are consistent between holes and confirm that the dynamics of hole growth on pure components ($\phi = \{0,1\}$) are slower than the dynamics on all mixed SAMs; this trend is consistent with the slip lengths reported in Figure~\ref{slipfig}. 

\section{\label{appB}Derived quantities from XRR fits}

\begin{figure}[t!]
\includegraphics[width=\columnwidth]{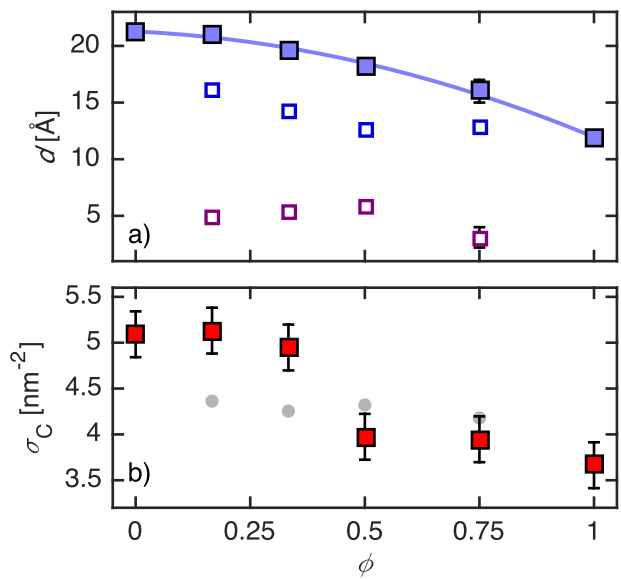}
\caption{\label{tailschains} a) Thickness of the SAM tail layers (filled blue) as a function of composition. For the bidisperse SAMs, $d = d_1+d_2$, with the components $d_1$ and $d_2$ shown as open squares (same color scheme as shown in Figure~\ref{EDPmods}; {\em cf.} also Table~\ref{XRRtable}). The solid line is a guide to the eye.  b) Grafting density of chains, computed according to Equation~\ref{graftedchains}; grey circles show $\srm{\sigma}{C}^\textrm{alternate}$. }
\end{figure}

In this section we provide supplementary derived quantities from the electron density profiles $\srm{\rho}{e}(z)$ presented in Figure~\ref{EDPmods}. First, we show the thickness of the tail layers $d$ in Figure~\ref{tailschains} (\emph{cf}. Table~\ref{XRRtable}), which, as noted in the main text and contrasting with the friction experienced by PS melts on these SAMs, is monotonic with the composition. 

The EDPs can also be used to compute various areal densities once the EDP of the tail layers is integrated. Using the models shown in Figure~\ref{EDPmods}, areal electron density of the silane tails is obtained through 
\begin{align}
	\srm{\sigma}{e} = \int dz\ \rho_\textrm{e}(z)\ , 
\label{electrondensity}
\end{align}
where the integration is taken over the tail layer(s) of the EDP (blue and violet components only). The grafting density of chains, \srm{\sigma}{C}, is then \srm{\sigma}{e}\ divided by the average number of electrons per grafted alkyl chain, \srm{N}{OTS}, for {\em e.g.} OTS. Thus, $\srm{\sigma}{C,OTS} = \srm{\sigma}{e,OTS}/\srm{N}{OTS}$ and likewise for DTS; accounting for the electrons associated with elemental C and H in the molecules, $\srm{N}{OTS} = 145$ and $\srm{N}{DTS} = 97$. 

For the mixed monolayers, we have provided above an estimate for the exposed chain density, \srm{\sigma}{ex}, Equation~\ref{exchainsmain}. Here we provide furthermore an estimate for the total chain grafting density along similar lines. By integrating the EDP for tail layer 1, the areal electron density, $\sigma_1$, of an effective layer of DTS chains is accounted for; this is the case since $\sigma_2$ has already accounted for the parts of OTS molecules with molar mass in excess of DTS. Thus, the total grafting density of alkyl chains in the SAM is approximated by
\begin{align}
	\srm{\sigma}{C} = 
	\begin{cases}
	    \srm{\sigma}{e}/\srm{N}{OTS}\ ,			& \text{if } \phi = 0\ ,\\
	    \srm{\sigma}{1}/\srm{N}{DTS}\ , 	& \textrm{if }0<\phi<1\ , \\
    	    \srm{\sigma}{e}/\srm{N}{DTS}\ ,			& \text{if } \phi = 1\ .
	\end{cases}
\label{graftedchains}
\end{align} 
It is also possible simply to assume that the volume fraction of DTS chains in the SAM is the same as that which was in the solution used to prepare the SAMs. In this case, Equation~\ref{electrondensity} can be used to give an alternate estimate for the density of grafted chains, given as
$\srm{\sigma}{C}^\textrm{alternate} = \srm{\sigma}{e}[{\srm{N}{OTS}(1-\phi)+\srm{N}{DTS}\phi}]^{-1}$. The chain grafting density is shown in Figure~\ref{tailschains}b). As with the exposed chain density, \srm{\sigma}{ex}, the general trend of \srm{\sigma}{C}\ with $\phi$ is common between the two estimates. We note in contrast to \srm{\sigma}{e}, however, that \srm{\sigma}{C}\ is monotonic with $\phi$.

\section{\label{appC}Supporting Molecular Dynamics Results}

MD simulations in the isobaric-isothermal ensemble~\cite{frenkel02TXT} (NPT) were performed for different DTS and OTS self- assembled monolayers on silicon oxide at 298 K and 1 bar with the Gromacs simulation package,~\cite{pronkBINF13} version 4.6.2. The supporting substrate was modeled as a flat b-cristobalite (1 0 1) surface (placed normal to the $z$-axis) with the dimensions ($11.5 \times 11.1$)\,nm$^2$ and thickness of about 2.3 nm. To obtain alkylsilane SAMs with different density, an appropriate number of DTS or OTS molecules was bonded to randomly selected oxygen atoms on the top side of the substrate. Every bonded molecule was initially placed perpendicular to the surface in a random orientation with an all-trans conformation. Periodic boundary conditions in all directions were used in the simulations. Details about the force field, molecular dynamic parameters and other simulation procedures can be found in ref.~\onlinecite{castillo15LAN}.

\begin{figure}[t!]
\includegraphics[width=\columnwidth]{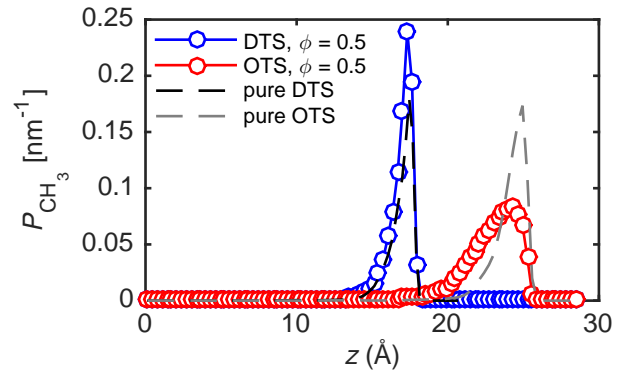}
\caption{\label{MDfig} Localization probability density of the terminal CH$_3$ groups in OTS and DTS molecules for a mixed monolayer with 50\% OTS (red) and 50\% DTS (blue), and for pure SAMs containing only OTS or DTS (dashed lines). The probability densities were computed using MD simulations as in ref.~\onlinecite{castillo15LAN}. The grafting density was $\srm{\sigma}{C} = 4.5$\,nm$^{-2}$ in all cases. }
\end{figure}

\section{\label{appD}Table of fitting parameters for XRR data}
\begin{table*}[t]
\caption{\label{XRRtable}Fitting paramaters of the slab models obtained using the effective density model~\cite{tolan99TXT} for the X-ray reflectivity data shown in Figure~3. Electron density $\rho_\textrm{e}$, roughness $\sigma$, and thickness $d$ are shown. Errors on the last digit are indicated in brackets, except where the parameters were not varied in the fit. The last row shows the total thickness of the tail layer, which for the mixed SAMs is the sum of thicknesses for Tail 1 and Tail 2 (blue and violet curves in Figure 4, respectively). }
\vspace{1mm}
\begin{tabular}{l l | c c @{\hskip 0.35in} c c c c @{\hskip 0.35in} c c}
\hhline{==========}
 & & \makecell{OTSQ\\($\phi$ = 0)} & \makecell{OTS\\($\phi$ = 0)} & \makecell{OTS/DTS\\($\phi$ = 0.17)} & \makecell{OTS/DTS\\($\phi$ = 0.33)} & \makecell{OTS/DTS\\($\phi$ = 0.5)} & \makecell{OTS/DTS\\($\phi$ = 0.75)} & \makecell{DTS\\($\phi$ = 1)} & \makecell{DTSQ\\($\phi$ = 1)} \\ 
\hline

\hline
\rule{0pt}{3ex}  
\multirow{2}*{Quarz} 
&$\rho_\textrm{e}$ [\AA$^{-3}$] 				& 0.875 & - & - & - & - & - & - & 0.858 \\ 
 & $\sigma$ [\AA] 				& 1.82(7) & - & - & - & - & - & - & 2.31(5) \\ 
\rule{0pt}{6ex}  

\multirow{2}*{Si} 
&$\rho_\textrm{e}$ [\AA$^{-3}$] & - & 0.702 & 0.713 & 0.702 & 0.702 & 0.713 & 0.702 & - \\ 
 & $\sigma$ [\AA] & - & 1.8(2) & 1.5(4) & 3.7(2) & 2.2(2) & 3.0(6) & 1.5(1) & - \\ 
\rule{0pt}{3ex}  

\multirow{3}*{aSiO$_2$} & $d$ [\AA] & - & 10.1(2) & 8.1(7) & 10.4(3) & 8.5(2) & 9.6(7) & 9.5(1) & - \\ 
&$\rho_\textrm{e}$ [\AA$^{-3}$] & - & 0.677(1) & 0.688(9) & 0.665(1) & 0.665(2) & 0.671(5) & 0.664(1) & - \\ 
 & $\sigma$ [\AA] & - & 1.3(6) & 2.8(3)& 1.71(3) & 1.6(1) & 1.8(2) & 2.73(5) & - \\ 
\rule{0pt}{6ex}  

\multirow{3}*{Head} & $d$ [\AA] & 6.6(4) & 5.75(2) & 4.7(5) & 6.26(3) & 6.98(4) & 6.94(8) & 6.22(5) & 6.2(4) \\ 
&$\rho_\textrm{e}$ [\AA$^{-3}$] & 0.558(30) & 0.538(1) & 0.508(9) & 0.452(1) & 0.520(1) & 0.498(5) & 0.460(1) & 0.509(20) \\ 
 & $\sigma$ [\AA] & 5.1(5) & 2.76(6) & 3.4(3) & 1.81(7) & 3.09(6) & 2.6(2) & 3.90(8) & 3.3(5) \\ 
\rule{0pt}{6ex}  

\multirow{3}*{Tail 1} & $d_1$ [\AA] & 21.6(3) & 21.20(3) & 16.2(2) & 14.32(4) & 12.53(4) & 12.8(1) & 11.85(4) & 12.0(4) \\ 
&$\rho_\textrm{e}$ [\AA$^{-3}$] & 0.343(20) & 0.351(1) & 0.309(4) & 0.308(1) & 0.305(1) & 0.290(2) & 0.292(2) & 0.325(10) \\ 
 & $\sigma$ [\AA] & 4.0(4) & 3.43(3) & 1.8(5) & 2.2(1) & 3.34(8) & 2.6(2) & 4.23(9) & 3.7(4) \\ 
\vspace{-2mm}& & & & & & & & & \\
\multirow{3}*{Tail 2} & $d_2$ [\AA] & - & - &4.8(3) & 5.35(3) & 5.71(8) & 3.1(9) & - & - \\ 
&$\rho_\textrm{e}$ [\AA$^{-3}$] & - & - & 0.21(2) & 0.202(2) & 0.142(3) & 0.22(5) & - & - \\ 
 & $\sigma$ [\AA] & - & - & 1.5(6) & 2.0(1) & 2.7(1) & 3.3(5) & - & - \\ 
\vspace{-2mm}& & & & & & & & & \\
Tail total & $d$ [\AA] & 21.6(3) & 21.20(3) &21.0(5) & 19.7(1) & 18.2(1) & 16(1) & 11.85(4) & 12.0(4) \\ 
\hhline{==========}
\end{tabular}
\end{table*}


\section*{references}

\end{document}